\documentclass[review]{elsarticle}
\usepackage{amsmath,latexsym}

\usepackage{hyperref}
\journal{Journal of \LaTeX\ Templates}

\usepackage{bm}
\usepackage{graphicx,color}
\usepackage{physics}
\usepackage{dsfont}
\usepackage[normalem]{ulem}
\usepackage{mathrsfs}
\usepackage{mathtools}
\usepackage{calrsfs}
\usepackage{float}
\usepackage{subfigure}
\usepackage{tabularx}
\usepackage{colortbl}
\usepackage{dcolumn}
\usepackage{array}
\usepackage{varioref}
\usepackage[active]{srcltx}

\expandafter\ifx\csname package@font\endcsname\relax\else
\expandafter\expandafter
\expandafter\usepackage
\expandafter\expandafter
\expandafter{\csname package@font\endcsname}%
\fi
\hypersetup{
	colorlinks=true,       % false: boxed links; true: colored links
	linkcolor=blue,          % color of internal links
	citecolor=blue,        % color of links to bibliography
}

\usepackage[section]{placeins}

\begin{document}

\begin{frontmatter}

\title{Maximal momentum GUP leads to quadratic gravity}

\author{Vijay Nenmeli\fnref{vijay.nenmeli@iitb.ac.in}}
\address{Department of Physics, Indian Institute of Technology Bombay, Mumbai 400076, India}
\author{S. Shankaranarayanan\fnref{Corresponding Author, shanki@phy.iitb.ac.in}}
%\email{shanki@phy.iitb.ac.in}
\address{Department of Physics, Indian Institute of Technology Bombay, Mumbai 400076, India}
\author{Vasil Todorinov\fnref{v.todorinov@uleth.ca}}
 \address{Theoretical Physics Group and Quantum Alberta, Department of Physics and Astronomy, University of Lethbridge, 4401 University Drive, Lethbridge, Alberta, T1K 3M4, Canada}
\author{Saurya Das\fnref{saurya.das@uleth.ca}}
\address{Theoretical Physics Group and Quantum Alberta, Department of Physics and Astronomy, University of Lethbridge, 4401 University Drive, Lethbridge, Alberta, T1K 3M4, Canada}

\begin{abstract}
Quantum theories of gravity predict interesting phenomenological features such as a minimum measurable length and maximum momentum.
We use the Generalized Uncertainty Principle (GUP),
which is an extension of the standard Heisenberg Uncertainty Principle motivated by Quantum Gravity, to model the above features. In particular, we use a GUP with modelling maximum momentum to establish a correspondence between the GUP-modified dynamics of a massless spin-2 field and quadratic (referred to as Stelle) gravity.
In other words, Stelle gravity can be regarded as the classical manifestation of a maximum momentum and the related GUP. We explore the applications of Stelle gravity to cosmology and specifically show that Stelle gravity applied to a homogeneous and isotropic background leads to inflation with an exit. 
Using the above, we obtain strong bounds on the GUP parameter from CMB observations. Unlike previous works, which fixed only upper bounds for GUP parameters, we obtain both \emph{lower and upper bounds} on the GUP parameter. 
\end{abstract}


\end{frontmatter}

%\linenumbers

A common feature found in most theories of Quantum Gravity (QG) is the existence of a minimum measurable length and/or a maximum measurable momentum of particles. This gives rise to a modification of the standard Heisenberg algebra and leads to the so-called Generalized Uncertainty Principle (GUP)\cite{Adler1999-db,Adler_2001,
Ali:2010yn,
Ali2011,
Ali_2014,
Ali_2015,
Alonso_Serrano_2018,Amati1989-gs,
%Amelino-Camelia1997-xx,
Amelino_Camelia_2001,
AMELINO_CAMELIA_2002,
Amelino-Camelia2013-xs,
Bargue_o_2015,Bambi2007-te,Bawaj_2015,
Bojowald2011-bb,Bolen2005-jq,
%Bonder:2017ckx,
%Bosso:2017hoq,
bosso2017generalized,
Bosso2018,
Bosso:2018uus,
Bosso:2019ljf,
%Bosso:2020aqm,
%Bosso:2020ztk,
Burger_2018,
bushev2019testing,
Capozziello:1999wx,
Casadio_2020,
Chang:2011jj,
Cortes:2004qn,
Costa_Filho2016-ox,
Dabrowski2019-cb,
Dabrowski2020-kk,
Das2008,
Das:2009hs,
Das:2010zf,
DAS2011596,
Das2014,
Das_2019,
Das_2020,
%das2021test,
%das2021bounds,
%Gangopadhyay_2018,
Garcia-Chung:2020zyq,
Giddings2020-xz,
Hamil2019-qh,
%tHooft:1996ziz,
Hossenfelder:2006cw,
Hossenfelder:2012jw,
Kempf1995-ka,
Kober:2010sj,KONISHI1990276,
%Lehnert2011-mi,
MAGGIORE199365,
Maggiore:1994,
Marin:2013pga,
Mead1966-xj,
Moradpour2021-jy,
Mureika2019-lf,
Myung_2007,
%Nozari_2008,
Park2008-uj,
%Pikovski_2012,
%Rovelli:1994ge,
Scardigli1999-ne,
Snyder:1946qz,
Sprenger_2011,
Sriramkumar:2006qt,
Stargen_2019,
tedesco2011fine,
wang2016solutions}. A covariant version of GUP was proposed recently which admits a frame independent/Lorentz invariant  minimum length or maximum momentum scale \cite{Quesne2006,Todorinov2018-xi,Bosso2020-dv,Bosso2020-fz}.

%If minimal length is indeed a consequence of quantum gravity, one question naturally arises: 
Since an additional scale in the Universe is a consequence of QG, one question naturally arises: 
what are its observational consequences? 
%of these theoretical models? 
Much effort is focused on indirect signatures in high-energy particle collisions or astrophysics~\cite{Amelino-Camelia1997-xx,Scardigli2014-wm}. To make testable predictions in the strong gravity regimes, such as in the early Universe and in the proximity of black-holes, one needs to look at the GUP modifications to General Relativity (GR). This leads to the following related questions: What kind of corrections will GUP introduce to GR, will the resulting effective field theory preserve general covariance, and what are its testable predictions?

In this work, we use a GUP model with maximum momentum to establish a correspondence between the GUP modified dynamics of a massless spin-2 field and a class of quadratic (also referred to as Stelle) gravity~\cite{Stelle1978-du,Stelle1977-rd,Noakes1983-fo}. Thus, we show that Stelle gravity can be regarded as the classical manifestation of the momentum cutoff at the QG scale with \emph{no malicious ghosts}~\cite{Tomboulis1984-bb,Antoniadis1986-hj}. We then apply the QG induced Stelle type theory to the early Universe and obtain \emph{strong bounds} on QG parameters.  Similar estimations have been done in a host of low-energy laboratory-based experiments and phenomena~\cite{Amelino-Camelia1997-xx,Scardigli2014-wm}. To our knowledge, this is the first time such a result has been obtained in a cosmological context, and it yields one of the most stringent allowed parameter ranges.

%Stelle gravity has another important use, which is its application to 
%cosmological inflation. The theory provides a mechanism of inflation. In this letter, we exploit this fact to estimate values of QG parameters that define the GUP. 
%While similar estimations have been done in a host of low energy laboratory-based experiments and phenomena, to our knowledge, this is one of the earliest works in which such a result is obtained in the cosmological context, and indeed gives one of the best-allowed ranges of the parameters.

%the maximum momentum uncertainty version of GUP implies modifications of the equations of motion of the graviton. The equations of motion can, in turn, be used to reconstruct the gravitational action and obtain QG modified Einstein equations. Remarkably, the modified equations and the corresponding action turns out to belong to the class of Stelle gravity~\cite{Stelle1978-du,Stelle1977-rd}. 
%It is known that the theory, in general, has ghosts and tachyonic instabilities, which, as we shall, can be removed by a suitable choice of parameters. 
%
%The theory has ghosts and therefore fails to serve as a consistent theory of QG.
%
%This milestone

%One of the crucial ingredients in this analysis is Feynman's approach to gravitation~\cite{Feynman2002-hb}. 
We use the {Gupta-Feynman} approach to gravitation to obtain an effective GUP corrected gravity theory as it derives the gravity equations of motion based on the underlying quantum aspects of gravity~\cite{Gupta1954, Feynman2002-hb}.
More specifically, {in this approach,} Einstein's general theory of relativity is derived as the inevitable result of the demand for a self-consistent theory of a massless spin-2 field (the graviton) coupled to the energy-momentum tensor of matter. One of the advantages of this approach is the intimate and fundamental connection between gauge invariance and the principle of equivalence~\cite{Gupta1954, Feynman2002-hb}. We first study the lowest-order Relativistic GUP (RGUP) effects on a massless spin-2 field in the Minkowski background.  We obtain the RGUP corrected gauge-invariant gravity action by demanding that the effective theory of gravity containing GUP correction be gauge-invariant. Interestingly, this minimal RGUP corrected modified gravity theory belongs to the class of \emph{Stelle gravity theory}~\cite{Stelle1977-rd,Stelle1978-du,Noakes1983-fo}.

%\Saurya{The above paragraph is fine, but my read better is shortened, without sacrificing content.} \Shanki{Trimmed a bit.}

Our starting point is the covariant GUP, recently introduced by Todorinov et al. ~\cite{Todorinov2018-xi}: 
 \begin{equation}
     [x^{\mu},p^{\nu}]=i\hbar\eta^{\mu\nu}(1 - \gamma p^{\sigma}p_{\sigma})-2i\hbar\gamma p^{\mu}p^{\nu},
     \label{def:GUPcomm}
 \end{equation}
where $\gamma=\gamma_0/(M_{\text{Pl}}c)^2$ is the Lorentz invariant scale.
 The reader can easily see that the physical position ($x^{\mu}$) and momentum operators  ($p^{\nu}$) are not canonically conjugate. 
 So we introduce two new vectors, $x^{\mu}_0$ and $p^{\nu}_0$ which are canonically conjugate:
\begin{equation}
\label{def:CanoXP}
p_0^{\mu} =  -i\hbar\frac{\partial}{\partial x_{0\,\mu}}, \quad
[x_0^{\mu},p_0^{\nu}] =  i\hbar\eta^{\mu\nu} \, ,
\end{equation}
where we choose the physical position ($x^\mu$) to coincide with the auxiliary one ($x_0^\mu$). In Ref. \cite{Todorinov2018-xi}, the GUP modified  momentum operator was shown to be:
\begin{equation}
\label{GUPp01}
p^{\mu} = {p_0^\mu}/\left(1 + \gamma \, p_0^{\rho} p_{0\rho} \right) \, .
\end{equation}
To linear order in $\gamma$, the physical momentum can be written as: 
\begin{equation}
\label{GUPp}
p^{\mu} \simeq
p_0^\mu\, \left(1 - \gamma \, p_0^{\rho} p_{0\rho} \right)\,.
\end{equation}
For the chosen representation of the physical position ($x^\mu$) and momentum ($p^\mu$), the dispersion relation and the irreducible representations of the modified Poincar\'e group are preserved and correspond to maximum momentum RGUP~\cite{Todorinov2018-xi}. Since the dispersion relations are preserved,  substituting Eq.\eqref{GUPp} into 
$p_{\mu} p^{\mu} = - m^2$, leads to the following modified Klein-Gordon equation~\cite{Bosso2020-dv,Bosso2020-fz}:
\begin{equation}\label{ModKG}
    -\Box\phi+2\gamma\Box^{2}\phi + m^{2}\phi=0\,,
\end{equation}
where we have set $\hbar = c = 1$. The Ostrogradsky 
method~\cite{Pons:1988tj,Woodard:2015zca,deUrries:1998obu} leads to the following modified Klein-Gordon Lagrangian:
\begin{equation}\label{KGfieldLagrangian}
\mathcal{L}_{\rm KG}^{\rm GUP} =\mathcal{L}_{\rm KG} + 
\gamma \, \partial^{\nu} \phi  \, 
\Box \left(\partial_{\nu} \phi\right) \, ,
\end{equation}
where $\mathcal{L}_{KG}$ denotes the unmodified Klein-Gordon Lagrangian. It was shown
that for energies below the Planck energy, effective action Eq.\eqref{KGfieldLagrangian} does not have Ostrogradsky ghosts as shown in Appendix E of ~\cite{Todorinov2020}. 

In the irreducible representations of the modified Poincar\'e group, there is a spin-2 particle consistent with the equation of motion Eq.~\eqref{ModKG}. 
We use this fact and apply the {Gupta-Feynman} approach to gravity~\cite{Gupta1954, Feynman2002-hb} to the spin-2 field and arrive at the RGUP corrected gravity action. 

The procedure begins by obtaining the RGUP corrections to the massless spin-2 fields in the flat background. The {Gupta-Feynman} approach allows us to obtain the quantum field theory describing massless spin-2 
($h_{\mu\nu}$) quanta (gravitons) coupled to matter in the Minkowski background, using two assumptions --- Lorentz invariance and the presence of only two helicity states for the massless spin-2 graviton~\cite{Gupta1954, Feynman2002-hb}. 
Thus, the Lagrangian of the massless spin-2 field that is linearly coupled to a conserved energy-momentum tensor is given by:
\begin{eqnarray}
\label{eq:action01}
%\begin{multlined}
\mathcal{L}_h &=& a \, \partial_{\sigma}h_{\mu\nu}\partial^{\sigma}h^{\mu\nu} 
+b \, \partial_{\mu}h^{\mu\nu}\partial_{\sigma} h^{\sigma}_{\nu}  \\
&  &
+ c \,  \partial_{\nu}h^{\mu\nu} \partial_{\mu}h^{\sigma}_{\sigma}   +d \, \partial_{\mu}h^{\nu}_{\nu} \partial^{\mu}h^{\sigma}_{\sigma}  
+  \lambda \, h_{\mu\nu}T^{\mu\nu}\,, \nonumber
%\end{multlined}
\end{eqnarray}
where $\lambda$ is the coupling constant and the constants $a, b, c, d$ are  determined by demanding the conservation of the energy-momentum tensor $\partial_{\mu} T^{\mu\nu} =0$.
%We get $a = -d = 1/2, c = -b = 1$.
%  
We demand that the theory is gauge-invariant and that the field equation satisfied by $h_{\mu\nu}$ be compatible with the matter equations of motion. %Feynman infers that
In that case, higher-order non-linear corrections must be added to the action in Eq.\eqref{eq:action01} leading to Einstein's equations. Additionally, we assume that the RGUP corrections are also gauge-invariant and that the field equation satisfied by $h_{\mu\nu}$ is compatible with the matter equations of motion. As we show, these two conditions uniquely lead to RGUP modifications to GR. 

Equations of motion of the massless spin-2 field corresponding to the action \eqref{eq:action01} are:
\begin{eqnarray}
\label{eq:EOM01} 
2 \, G_{\mu\nu}^{(L)} \equiv & & -  
\left[ \partial^{\sigma} \partial_{\sigma} h_{\mu\nu}
- \partial^{\sigma} \partial_{\nu} h_{\mu\sigma}
- \partial^{\sigma} \partial_{\mu} h_{\nu\sigma} \right. \\
& & - \left. \partial_{\mu} \partial_{\nu} h^{\sigma}_{\sigma}
+\eta_{\mu\nu} \partial^{\sigma} \partial_{\rho} h^{\sigma\rho}
-\eta_{\mu\nu} \partial^{\rho} \partial_{\rho} h^{\sigma}_{\sigma} \right] 
=- \lambda T_{\mu\nu} \, , \nonumber
\end{eqnarray}
where $G_{\mu\nu}^{(L)}$ is the linearized Einstein tensor. Following Eq. \eqref{GUPp}, substituting $\partial_{\mu}\rightarrow \partial_{\mu}(1- \gamma \Box)$, the RGUP modified equations of motion for the massless spin-2 field in the Minkowski background become:
\begin{equation}
\label{eq:EOM02} 
%\partial^{\sigma} \partial_{\sigma} h_{\mu\nu} 
%- \partial^{\sigma} \partial_{\nu} h_{\mu\sigma} 
%- \partial^{\sigma} \partial_{\mu} h_{\nu\sigma} 
%- \partial_{\mu} \partial_{\nu} h^{\sigma}_{\sigma}
%+\eta_{\mu\nu} \partial^{\sigma} \partial_{\rho} h^{\sigma\rho}
%-\eta_{\mu\nu} \partial^{\rho} \partial_{\rho} h^{\sigma}_{\sigma} 
G_{\mu\nu}^{(L)} 
+ \gamma \, {\cal  G}_{\mu\nu}
=-\lambda T_{\mu\nu}/2 \,,
\end{equation}
where 
\begin{eqnarray}
{\cal G}_{\mu\nu} &=& \Box^{2}h_{\mu\nu} 
- \Box  \left( \partial^{\sigma} \partial_{\nu} \right) h_{\mu\sigma} 
- \Box \left( \partial^{\sigma} \partial_{\mu} \right) h_{\nu\sigma}  \\
& & 
+ \Box \left( \partial_{\mu} \partial_{\nu} \right)  h^{\sigma}_{\sigma} + \eta_{\mu\nu}\Box \left( \partial_{\sigma} \partial_{\rho} \right)
h^{\sigma\rho} -\eta_{\mu\nu}\Box^{2}h^{\sigma}_{\sigma}\,. \nonumber
\end{eqnarray}
The LHS of Eq.~\eqref{eq:EOM02} can be derived from the following Lagrangian:
\begin{eqnarray}
\mathcal{L}_{h}^{\rm GUP} &=&\mathcal{L}_{h} 
-  \gamma \left[ 
\partial^{\sigma} h^{\mu\nu} \, \Box \left( \partial_{\sigma} h_{\mu\nu} \right)  - 2 \partial^{\sigma} h_{\sigma\nu}  \Box \left( \partial_{\mu}  h^{\mu\nu} \right) \right. \nonumber \\
\label{eq:action02}
& & \left. 
 + 2 \partial_{\mu} h^{\sigma}_{\sigma} \Box \left( \partial_{\nu}  h^{\mu\nu} \right) -  \partial^{\mu} h^{\sigma}_{\sigma}  \Box \left( \partial_{\mu} h^{\nu}_{\nu} \right) \right]\,. 
\end{eqnarray}
%
%This is the GUP corrected Lagrangian corresponding to the massless spin-2 field without imposing gauge invariance. 
{This linear higher-derivative theory contains ghosts. However, as was 
shown in Ref. \cite{Antoniadis1986-hj}, the  non-linear theory does not contain malicious ghosts.} The next step is to impose gauge-invariance in the above action \eqref{eq:action02}. We demand that the equations of motion remain invariant under the transformation:
%
%\begin{equation}
%\label{eq:GaugeCond01}
$     h'_{\mu\nu} \rightarrow h_{\mu\nu}+{X^{{(S)}}_{\mu,\nu}} ,
$
%\end{equation}
%
where $X_{\mu}$ is an arbitrary vector field, and the superscript $(S)$ denotes the operation of symmetrization. Thus, the solutions to Eq.~\eqref{eq:EOM02} are not unique and we need to pick a particular solution from the large list of possibilities. We use the \emph{Lorenz gauge} condition, i. e.,
%
%\begin{equation}
%\label{eq:GaugeCond02}
$\overline{h^{\mu\nu}_{,\nu}}=0$, where, 
$\overline{X_{\mu\nu}}=X^{{(S)}}_{\mu\nu} - \eta_{\mu\nu}X^{\rho}_{\rho}/2$ \, ,
the overbar denotes the combined operations of symmetrization and trace reversing. In this gauge, the source-free equation \eqref{eq:EOM02} reduces to:
\begin{equation}
\label{eq:EOM03} 
\Box h_{\mu\nu}-2\gamma\Box^{2}h_{\mu\nu}=0\,.
\end{equation}
We would like to draw attention to the following points for this first key result: First, in the limit of $\gamma \to 0$, the above expression leads to the well-known wave-equation in linearized gravity that defines gravitational radiation. 
In this limit, the spin-2 graviton has only two helicity states~\cite{Gupta1954, Feynman2002-hb}.
Second, the above RGUP modified wave-equation leads to a classical theory with more than two helicity states. To see this, substituting the plane-wave ansatz ~$h_{\mu\nu}(x)=e_{\mu\nu}e^{ik.x}$ in Eq.~\eqref{eq:EOM03}, leads to:
%
% \begin{equation}
% \label{eq:Pol01} 
$k^{2}(1+2\gamma k^{2})=0$, and   $e_{\mu\nu}k^{\mu}=0$.
This leads to two class of solutions ---
$k^{2}=0~\mbox{or}~k^{2}= - {1}/(2\gamma). 
$
This implies that besides the massless spin-2 mode (corresponding to $k^{2}=0$), there exists a massive spin-2 mode with mass $\sqrt{1/2 \gamma}$. 
%
%\begin{equation}
%\label{eq:Propogator}
%      P_{\mu\nu\sigma\tau}=\frac{\eta_{\mu\sigma}\eta_{\nu\tau}+\eta_{\mu\tau}\eta_{\nu\tau}-\eta_{\mu\nu}\eta_{\sigma\tau}}{k^{2}(1+2\gamma k^{2})}\,.
%\end{equation}
% 
%From the denominator, it is clear that the RGUP modified theory leads to two propagators --- one corresponding to the usual massless graviton and the other corresponding to a massive spin-2 boson of mass $\sqrt{\frac{1}{2\gamma}}$. 
For the massless graviton mode, the condition $e_{\mu\nu}k^{\mu}=0$ 
and gauge-transformation 
$e_{\mu\nu}\rightarrow e_{\mu\nu}+k_{\mu}a_{\nu}+k_{\nu}a_{\mu}$ ($a_{\mu}$ is arbitrary), 
reduces the number of independent components of $e_{\mu\nu}$ from $10$ to $2$. For the massive spin-2 mode (corresponding to $k^{2}=- 1/(2\gamma)$), 
$e_{\mu\nu}k^{\mu}=0$ holds, while, the gauge-transformation is not satisfied. Hence, the RGUP corrected linearized equations \eqref{eq:EOM03} contain additional degrees of freedom.  This also implies that the RGUP parameter $\gamma > 0$ is positive. As we will show later, this plays an important role in making the theory well defined.

We obtain the RGUP corrections beyond the linear regime by using a more direct approach. As noted above, the RGUP corrected spin-2 Eq.~\eqref{eq:EOM03} contains fourth-order derivative terms. Thus, the classical gravity action must contain higher-order Riemann, Ricci tensor, and Ricci scalar terms. We therefore consider the following minimal action that contains second-order Ricci scalar and tensor terms, i. e.,
\begin{equation}
\label{eq:StelleAct}
S=\int \, d^4x \, \sqrt{-g} \left[ 
\frac{1}{2\kappa^{2}}R  - \alpha \, R^{\mu\nu}R_{\mu\nu} + \beta \, R^{2} 
\right] \, ,
 \end{equation}
where $\kappa^{2}=8\pi G$. The above action is referred to as Stelle gravity action~\cite{Stelle1977-rd}. Stelle gravity is a two-parameter theory; 
$\alpha$ and $\beta$ are undetermined and uncorrelated. Using linear 
analysis of the isotropic solutions of the above action, Stelle showed that 
in addition to the usual $1/r$ effective potential, the above action also contains two Yukawa solutions with Yukawa masses \textcolor{blue}{are} inversely proportional to $\sqrt{\alpha}$ and 
$\sqrt{3\beta-\alpha}$~\cite{Stelle1977-rd}. Avoiding tachyonic instabilities 
leads to the following conditions on the parameters:
  \begin{equation}
  \label{eq:StelleCond01}
        \alpha\geq 0~~{\rm and}~~3\beta\geq\alpha  \Longrightarrow \beta \geq 0 \,.
\end{equation}
%
%Thus, the above conditions imply that both the parameters must be positive. 

The variation of action \eqref{eq:StelleAct} w.r.t metric leads to~\footnote{
%The equations in Ref. \cite{Stelle1977-rd} use the opposite convention for the Riemann tensor as opposed to the convention used in this work. 
We follow the Riemann tensor definition in Ref. \cite{MTW}. The equations in Ref. \cite{Stelle1977-rd} use the opposite convention for the Riemann tensor as opposed to this work.}:
%Hence, there are signature differences for a few terms when comparing with the corresponding equation in Ref. \cite{Stelle1977-rd}.}: 
%
{\small
\begin{eqnarray}
\label{eq:StelleEOM01}
& & \frac{1}{\kappa^2} G_{\mu\nu} + 2(\alpha-2\beta) \nabla_{\mu} \nabla_{\nu} R 
-2 \alpha \Box R_{\mu\nu} -(\alpha- 4 \beta)g_{\mu\nu}\Box R  \nonumber \\
&-& 4 \alpha R^{\rho\sigma}R_{\mu\rho\nu\sigma} + 
4\beta RR_{\mu\nu}+ g_{\mu\nu}(\alpha R^{\rho\sigma}R_{\rho\sigma}-\beta R^{2}) = T_{\mu\nu}.~~~~~
\end{eqnarray}
}
Linearizing the Stelle equation about the Minkowski background leads to the following equation:
{\small
\begin{eqnarray}
\label{eq:StelleEOM02}
 & & G^{(L)}_{\mu\nu} 
- 2 \kappa^2 (\alpha-2\beta) 
\left[
\Box \left( \partial_{\mu} \partial_{\nu} \right)  h^{\lambda}_{\lambda} 
- \left( \partial^{\mu} \partial_{\nu} \partial^{\rho} \partial_{\lambda} \right)  
h^{\lambda}_{\rho} \right.   \\
& & - \left. \eta_{\mu\nu} (\Box^{2} h^{\lambda}_{\lambda}
-\Box \left( \partial_{\lambda} \partial_{\rho} \right)  
h^{\lambda\rho}) \right]  
+ \alpha \kappa^2 \left[ 
\Box^{2} h_{\mu\nu} + 
\Box \left( \partial_{\mu} \partial_{\nu} \right)  h^{\lambda}_{\lambda} \right.  
 \nonumber \\
& & - \left. \Box \left( \partial_{\lambda} \partial_{\nu} \right)  h^{\lambda}_{\mu}
- \Box \left( \partial_{\lambda} \partial_{\mu} \right)  h^{\lambda}_{\nu}
-\eta_{\mu\nu}(\Box^{2} h^{\lambda}_{\lambda}
-\Box \left( \partial_{\lambda} \partial_{\rho} \right)  h^{\lambda\rho}) \right] 
= 0 \,.  \nonumber
\end{eqnarray}
}
It is interesting to note that the above Eq.~\eqref{eq:StelleEOM02} resembles the RGUP modified 
spin-2 field in Minkowski background \eqref{eq:EOM02}. 
In particular, setting $\alpha=2\beta$
%
% \begin{equation}
% \label{eq:Stellecons01}
%      \alpha=2 \beta
% \end{equation}
 %
 in Eq.\eqref{eq:StelleEOM02} leads to:
 \begin{eqnarray}
  \label{eq:StelleEOM03}
 G_{\mu\nu} &+& \alpha \kappa^2 \left[ \Box^{2}h_{\mu\nu} 
 -\Box \left( \partial_{\sigma} \partial_{\nu} \right)   h^{\sigma}_{\mu} 
 -\Box \left( \partial_{\sigma} \partial_{\mu} \right)   h^{\sigma}_{\nu} \right.~~~\\
\nonumber &+& \left.\Box \left( \partial_{\mu} \partial_{\nu} \right)   h^{\sigma}_{\sigma} 
 +\eta_{\mu\nu}\Box \left( \partial_{\sigma} \partial_{\rho} \right) 
 h^{\sigma\rho} -\eta_{\mu\nu}\Box^{2}h^{\sigma}_{\sigma} \right] =0 \,.  
 \end{eqnarray}
The above equation is identical to \eqref{eq:EOM02} when
\begin{equation}
 \label{eq:Stellecons02}
\alpha= {\gamma}/{\kappa^{2}} ~ \rm{and} ~\beta = {\gamma}/{(2 \kappa^{2})} \, .
\end{equation}
This is the second key result of this work regarding which we want to stress the following points: First, we have established a correspondence between Stelle gravity and RGUP corrected massless spin-2 fields.  The two unknown parameters of Stelle gravity are related to $\gamma$ (GUP parameter which has dimensions of $[L]^2$). Second, the {constraints} \eqref{eq:Stellecons02} {satisfy} the inequalities \eqref{eq:StelleCond01}. This implies that the class of Stelle gravity theories with $\alpha=2\beta$ is a classical realization of the QG phenomenon of maximal momentum.
Interestingly, for this class of Stelle theories, the masses of the two additional Yukawa bosons --- 
$1/\sqrt{2\alpha\kappa^2}$ and 
$1/\sqrt{4(3\beta-\alpha)\kappa^2}$ --- \emph{coincide} $(m_{\text{Yukawa}}=1/\sqrt{2 \gamma})$. 
%Interestingly, the two additional massive gauge bosons that Stelle theory predicts have equal masses ($1/\sqrt{2 \gamma}$). 
This implies that the RGUP corrections to GR lead to Stelle gravity that possesses an inherent degeneracy.
Lastly, Stelle gravity action \eqref{eq:StelleAct} contains higher derivative terms which may lead to ghosts. However, in Ref. \cite{Antoniadis1986-hj}, the authors showed that Stelle gravity 
with $\alpha > 0$ does not have malicious ghosts. 
%From Eq.\eqref{eq:Stellecons02}, we see that for RGUP modified gravity, $\alpha \propto \gamma > 0$.
Hence,  ghosts do not pose problems for RGUP-modified gravity.

So far, we have obtained a minimal gravity model that 
contains the RGUP corrections of \cite{Todorinov2018-xi,Bosso:2020aqm,Bosso:2020ztk}
in the field theory limit. 
We showed that this minimal gravity model, with one unknown parameter $(\gamma)$, is a specific form of \emph{Stelle gravity}. 
In the rest of this work, we will 
constrain the RGUP parameter by looking at 
the implications in the early Universe.

Although cosmological inflation is highly successful in
explaining CMB observations, there is no unique mechanism that leads to inflation. Despite the simplicity of the inflationary paradigm, obtaining a canonical scalar field-driven inflationary model within GR requires a highly fine-tuned potential~\cite{1999-Lyth.Riotto-PRep,2006-Bassett.etal-RMP}. While $f(R)$ models are the simplest extensions of GR, they can lead to a period of accelerating expansion early in the Universe's history~\cite{1980-Starobinsky-PLB,2010-DeFelice.Tsujikawa-LRR,Myrzakulov:2014hca,2019-Johnson.etal-GRG}. {Since the Starobinsky model is a specific instance of Stelle gravity} [$\alpha=0$ in Eq. \eqref{eq:StelleAct}], one naturally wonders whether Stelle Gravity (and RGUP modified gravity in particular) will lead to similar dynamics. To see that this is indeed the case, we substitute the flat space FRW metric:
\begin{equation}
\label{eq:FRWmetric}
    ds^{2}=-dt^{2}+a^2(t)(dx^{2}+dy^{2}+dz^{2})\,.
 \end{equation}
in Eq.~\eqref{eq:StelleEOM01}. For $\alpha = 2\beta = \gamma/\kappa^{2}$, we get:
 \begin{eqnarray}
 \label{eq:TimeFriedman}
H^{2}-2\gamma \left[\dot{H}^{2} 
-2 H \, \ddot{H} - 6H^{2}\dot{H}\right] &= 
0~~~~ \\
\label{eq:SpaceFriedman}
 3H^{2}+2\dot{H}+\gamma \left[ 18H^{2}\dot{H} + 9{\dot{H}}^{2}+12 H \ddot{H}+2\dddot{H}\right] &= 0,~~~~
 \end{eqnarray}
where $H \equiv H(t) = \dot{a}(t)/a(t)$ is the Hubble parameter. The above two equations are not independent and are related via the Bianchi identities. Specifically, adding Eq. \eqref{eq:TimeFriedman} to thrice its derivative leads to Eq.~\eqref{eq:SpaceFriedman}. Thus, we only solve Eq. \eqref{eq:TimeFriedman}.

It is interesting to note the similarities between Eqs. \eqref{eq:TimeFriedman} and \eqref{eq:SpaceFriedman}
 above and the EOM obtained from variation of the Starobinsky action \cite{1980-Starobinsky-PLB,2010-DeFelice.Tsujikawa-LRR,2019-Johnson.etal-GRG}. (Details in Appendix)
Therefore, we can conclude that RGUP modified gravity can be \emph{mapped} to Starobinsky gravity in the sense that the dynamics predicted by both models are the same. Hence, RGUP modified gravity (and generic Stelle gravity theory for that matter) will lead to inflation with exit.

The non-linear nature of the above equation means that parameter estimation can only be carried out numerically and by using cosmological observations. Note that the small value of the RGUP parameter ($\gamma$) and the large values of the Hubble parameter $H(t)$ make it impractical to solve Eq.~\eqref{eq:TimeFriedman} numerically.
\begin{figure}[!htb]
 \begin{flushleft}
  \begin{minipage}{0.5\textwidth}
    \begin{flushleft}
    \begin{table}[H]
    \begin{center}
    \begin{tabular}{c c} 
    \hline
         U~&$s_{0}$\\
         \hline
         ~~~$10^{-4}$~~~&$7.9*10^{-4}$\\
         ~~~$10^{-3}$~~~&$4.43*10^{-3}$\\ 
         ~~~$10^{-2}$~~~&$4.08*10^{-2}$\\ 
         ~~~$10^{-1}$~~~&$4.18*10^{-1}$\\
         ~~~$1$~~~&$9.5$\\
          ~~~$10$~~~&$1.18*10^{2}$\\
           ~~~$10^{2}$~~~&$1.21*10^{3}$\\
            ~~~$10^{3}$~~~&$1.21*10^{4}$\\
             ~~~$10^{4}$~~~&$1.21*10^{5}$\\
         \hline\hline
    \end{tabular}
    \end{center}
\end{table}
\end{flushleft}
  \end{minipage}
  \begin{minipage}[c]{0.35\textwidth}
\begin{flushleft}
    \includegraphics[height=\textwidth]{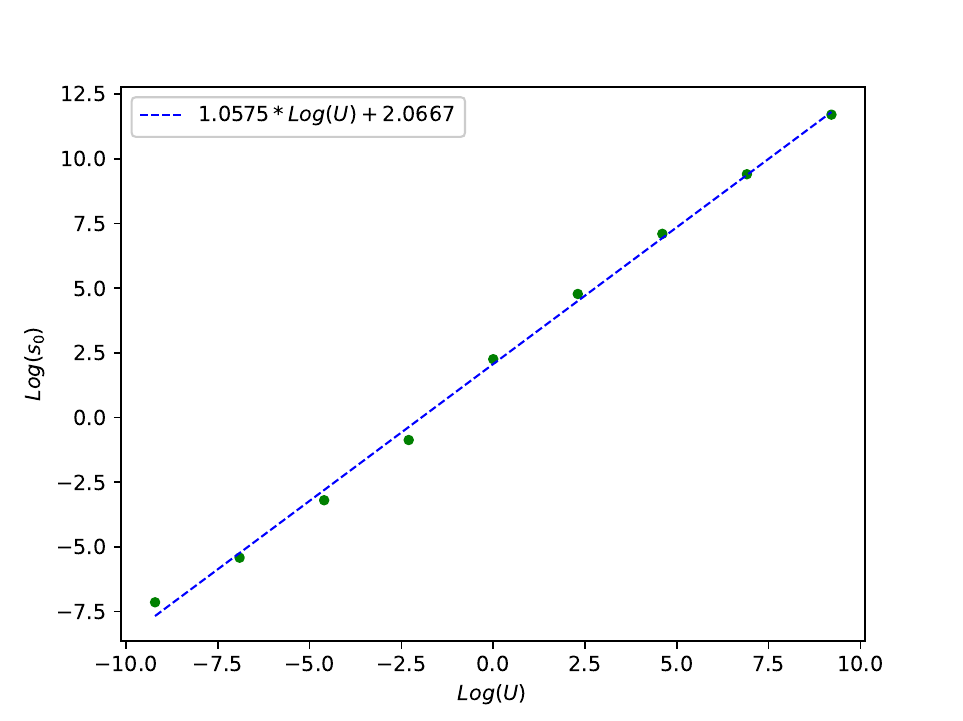}
    \end{flushleft}
  \end{minipage}
  \end{flushleft}
  \caption{The table shows the exit time $s_0$ for different values of $U$. The plot shows the linear correlation between the logarithms of $(U, s_0)$ supporting the $s_0\sim U$ conjecture.}\label{Fig:1}
\end{figure}

In order to make Eq. \eqref{eq:TimeFriedman} more manageable, we define the rescaled cosmic time $s$ as $s=t/\sqrt{\gamma}$. We also define $b(s)= \ln(H(s)/H_{\rm Inf})$ and 
$U = \sqrt{\gamma} \, H_{\rm Inf}$,
where $H(s)$ is the Hubble parameter w.r.t the dimensionless time $s$ and $H_{\rm Inf}$ is the inflationary energy scale. Physically, $U$ is the ratio of inflationary and RGUP energy scales. The slow-roll parameter $(\epsilon)$ is $\epsilon=(b'(s)e^{-b(s)})/U$.  In terms of these dimensionless variables, Eq.~\eqref{eq:TimeFriedman} becomes:
\begin{equation}
\label{eq:TimeFriedman02}
1 +2 \dot{b}^{2}(s) +4 \ddot{b}(s) +12 \, U \, e^{b(s)} \, \dot{b}(s)=0\,.
\end{equation}

For a range of $U$, we see that inflation ($\epsilon\leq 1$) occurs, and the exit occurs (See Appendix). The table in Fig. \ref{Fig:1} lists the exit time $(t_0, s_{0})$ 
(when $\epsilon=1$) as a function of $U$.
From Fig. \ref{Fig:1} we infer that $s_{0}\sim U$ for all values of $U$ considered. Assuming that such a relation holds generally, we now obtain bounds on the RGUP parameter $\gamma$. 

To do this, we define the following dimensionless parameters:
$
\gamma_0=\gamma/l_{Pl}^{2}, 
\alpha = t_0/l_{Pl}
$, 
where $l_{Pl}$ is the Planck length. Thus, the exit time $(s_0)$, can be rewritten as:
\begin{equation}\label{eq:exittime}
     s_{0}={\alpha \, l_{Pl}}/{\sqrt{\gamma}}= {\alpha}/{\sqrt{\gamma_{0}}}\,.
\end{equation}
Using the fact that $s_0 \propto U$, the above equation leads to:
\begin{equation}\label{eq:gamma_0}
\gamma_{0} \sim {\alpha}/{(H_{\rm Inf} \, l_{Pl})}\,.
\end{equation}
$\gamma_0$ can now be constrained once we know $\alpha$ and $H_{\rm Inf}$. 
PLANCK 7-year data gives an upper bound on the inflationary energy scale to be $10^{-4}M_{Pl}$~\cite{Akrami2020-tq}. We choose 
$H_{\rm Inf} = 10^{-5}M_{Pl}$ (GUT energy scale). Generic inflationary models require $\alpha$ in the range $10^{6}-10^{12}$~\cite{Bassett:2005xm}. Substituting these in Eq.~\eqref{eq:gamma_0}, we get $10^{10}\leq\gamma_{0}\leq 10^{16}$. 

Since the Physics near inflationary energy scales is not well understood, we can obtain a refined estimate by noting that the energy scale of RGUP effects must be above $10~TeV$ (LHC scale). This correspond to $\gamma_{0}=10^{15}$. Thus, we obtain the following bound on {$\gamma$}:
\begin{equation}
10^{10}\leq\gamma_{0}\leq 10^{15} 
~\Longrightarrow~
10^{-28}GeV^{-2}\leq\gamma\leq 10^{-23}GeV^{-2}\,.
\end{equation} 
These are among the strictest bounds obtained for RGUP parameters. Interestingly, in addition to providing an upper bound (as is usually done for RGUP parameters), we also obtain a \textit{lower} bound on the RGUP parameter. 

In conclusion, we have explicitly shown that the relativistic extension of GUP, in particular the one describing a maximal momentum, leads to a specific form of Stelle gravity. Note that Stelle gravity is the lowest-order correction. It is important to note that, although it contains higher derivatives, the action we derived does not suffer from malicious ghosts or instabilities of the ground state~\cite{Tomboulis1984-bb}. In addition, the resulting theory is renormalizable. To our knowledge, this is the first time such a parallel has been drawn. Furthermore, for FRW background, we showed that the Stelle gravity action could be mapped to the Starobinsky inflationary model. Hence, the RGUP action leads to inflation with exit. Using the 
correlation between the exit times $s_0$, the invariant scale $\gamma=\gamma_0/(M_{\text{Pl}}c)^2$, and the energy scale of inflation $H_{\rm Inf}$, we obtained one of the tightest bounds on the invariant scale $\gamma$. 
Importantly, here we have obtained a concrete range for the scale, whereas in most other works, one has obtained only an upper bound, with the lower bound on $\gamma_0$ implicitly assumed to be one (so as not to cross over to trans-Planckian regimes). 

{Our approach has two caveats: First, the procedure followed in this work may not produce a unique, effective action at higher orders~\cite{Rubio2019,Latosh2021}. Hence, we may have to use alternative approaches to include higher-order RGUP corrections. Second, the GUP modified gravity action derived in this work is an effective description and ignores other interacting terms. At non-perturbative regimes, we need to include the interacting terms.}

RGUP modified gravity is degenerate --- the masses of the two additional gauge bosons are equal. The origin and consequences of this degeneracy are unclear and are currently under investigation. In the future, we intend to more accurately model the observables of inflation, namely, the spectral tilt $n_s$  and the tensor-scalar ratio $r$ in hopes of further improving the bounds on $\gamma$. In addition, we can use Stelle gravity in the derivation of QG effects on gravitational wave detection. With the launch of LISA approaching, this direction will be of particular interest in the future.  

\noindent {\bf Acknowledgments:} The authors thank S. Mahesh Chandran and Manu Srivastava for comments on the earlier version of the manuscript.  SS thanks Namit Mahajan for discussions regarding Refs. \cite{Tomboulis1984-bb}.
This work is part of the Undergraduate project of VN. The work of SS is supported by the SERB-MATRICS grant. This work is supported by the Natural Sciences and Engineering Research Council of Canada. The authors thank the service personnel in India and Canada whose untiring work allowed the authors to complete this work during the pandemic.
\vspace*{-10pt}

\section*{Appendix}

The equations of motion for generic Stelle gravity (i.e., with $\alpha$ and $\beta$ unconstrained) are
{\small
\begin{eqnarray}
\label{app:StelleEOM01}
 \frac{1}{\kappa^2} G_{\mu\nu}&+& 2(\alpha-2\beta) \nabla_{\mu} \nabla_{\nu} R 
-2 \alpha \Box R_{\mu\nu} -(\alpha- 4 \beta)g_{\mu\nu}\Box R   \nonumber \\
%%%%
 &-&4 \alpha R^{\rho\sigma}R_{\mu\rho\nu\sigma} + 
4\beta RR_{\mu\nu}+ g_{\mu\nu}(\alpha R^{\rho\sigma}R_{\rho\sigma}-\beta R^{2})  = T_{\mu\nu}\,.
\end{eqnarray}
}
Substituting the FRW metric \eqref{eq:FRWmetric} into these equations, we get: 
{\small
 \begin{eqnarray}
 \label{eq:TimeFriedmanGenStelle}
H^{2}-(12\beta-4\alpha)\kappa^{2}\left(\dot{H}^{2} 
-2 H \, \ddot{H} - 6H^{2}\dot{H}\right) = 0 & & ~~\\
\label{eq:SpaceFriedmanGenStelle}
 3H^{2}+2\dot{H}+ [6\beta-2\alpha] \kappa^{2} \left[ 18H^{2}\dot{H} + 9{\dot{H}}^{2}+12 H \ddot{H}+2\dddot{H}\right]= 0 & & ~~
 \end{eqnarray}
 }
 Eqs.~\eqref{eq:TimeFriedmanGenStelle} and \eqref{eq:SpaceFriedmanGenStelle} above represent a family of differential equations, which describe the dynamics of Stelle theory of gravity for FRW metric.  Within the conditions set for $\alpha$ and $\beta$ in Eq. (13), two members of this family of differential equations and their solutions will differ only by a numerical constant.
Therefore, the 
\textit{qualitative dynamics predicted by a Stelle action is independent of the chosen parameter values}. As a result, we can conclude that the qualitative dynamics of \textit{any} Stelle theory can be studied by observing the dynamics of a particular case. The specific case considered is $\alpha=0$. Defining $M>0$ to be $1/\sqrt{12\beta\kappa^{2}}$, we get the corresponding Stelle action as 
\begin{figure*}[!htb]
%    \centering
    \includegraphics[scale=0.45]{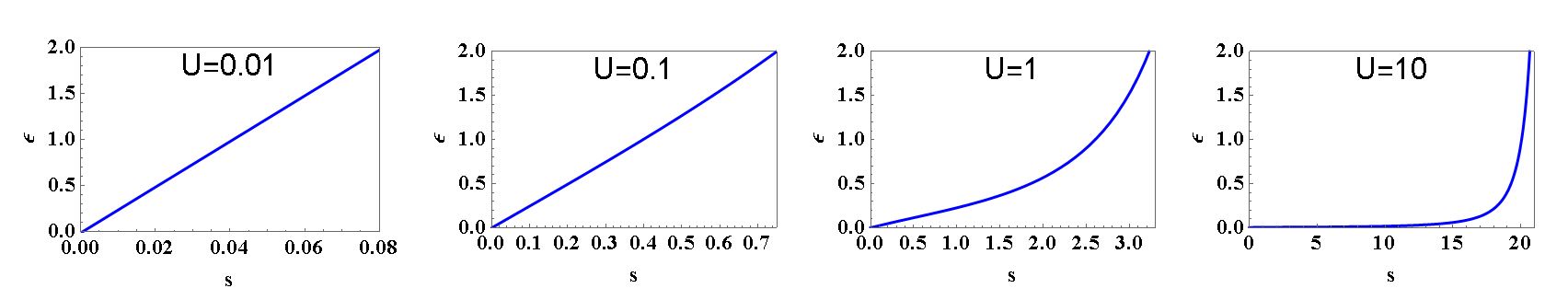}
    \caption{Plot of the $\epsilon$ vs $s$ for different values of $U$. We see that for a range of $U$, the model leads to inflation with exit.}
    \label{fig:2}
\end{figure*}
\begin{equation}
    S_{\alpha=0}=\frac{1}{2\kappa^{2}}\int{\sqrt{-g} \left(R+\frac{1}{6 M^{2}}R^{2} \right)d^{4}x}\,,
\end{equation}
which is the well-known Starobinsky action ~\cite{1980-Starobinsky-PLB}. The Starobinsky action leads to inflation with exit. From the preceding discussions, it is clear that any other Stelle theory (and GUP modified gravity in particular) will also lead to inflation with exit.  
To verify the analytic proof in the previous paragraph, we simulate Eq.~(21) and plot the slow-roll parameter $\epsilon$ as a function of the rescaled cosmic time $s_{0}$ for various $U$ values. From the plots of Figure 2, it is clear that $\epsilon$ increases monotonically from $0$ and crosses $1$ at finite times for all values of $U$. Thus, inflation with an exit is observed for all values of $U$ considered for a generic Stelle gravity.

%\bibliography{GUPRef}

\end{document}